\documentclass[fleqn,usenatbib]{mnras}

\usepackage{newtxtext,newtxmath}
\usepackage[T1]{fontenc}
\usepackage{ae,aecompl}

\usepackage{graphicx}	% Including figure files
\usepackage{amsmath}	% Advanced maths commands
\usepackage{amssymb}	% Extra maths symbols
\usepackage{physics}
\title[UWL pulsar polarimetry]{Pulsar polarimetry with the Parkes Ultra-Wideband receiver}

\author[Oswald et al.]{
Lucy Oswald$^{1}$\thanks{E-mail: lucy.oswald@physics.ox.ac.uk (LSO)},
Aris Karastergiou$^{1,2}$,
Simon Johnston$^3$
\\
$^{1}$Department of Astrophysics, University of Oxford, Denys Wilkinson Building, Keble Road, Oxford OX1 3RH, UK\\
$^{2}$Department of Physics and Electronics, Rhodes University, PO Box 94, Grahamstown 6140, South Africa\\
$^{3}$CSIRO Astronomy and Space Science, Australia Telescope National Facility, PO Box 76, Epping, NSW 1710, Australia \\
}

\date{Accepted XXX. Received YYY; in original form ZZZ}

\pubyear{2020}

\begin{document}
\label{firstpage}
\pagerange{\pageref{firstpage}--\pageref{lastpage}}
\maketitle

\begin{abstract}
Pulsar radio emission and its polarization are observed to evolve with frequency. This frequency dependence is key to the emission mechanism and the structure of the radio beam. With the new Ultra-Wideband receiver (UWL) on the Parkes radio telescope we are able, for the first time, to observe how pulsar profiles evolve over a broad continuous bandwidth of 700--4000 MHz. We describe here a technique for processing broadband polarimetric observations to establish a meaningful alignment and visualize the data across the band. We apply this to observations of PSRs~J1056--6258 and J1359--6038, chosen due to previously unresolved questions about the frequency evolution of their emission. Application of our technique reveals that it is possible to align the polarization position angle (PA) across a broad frequency range when constrained to applying only corrections for dispersion and Faraday rotation to do so. However, this does not correspond to aligned intensity profiles for these two sources. We find that it is possible to convert these misalignments into emission height range estimates that are consistent with published and simulated values, suggesting that they can be attributed to relativistic effects in the magnetosphere. We discuss this work in the context of the radio beam structure and prepare the ground for a wider study of pulsar emission using broadband polarimetric data. 
\end{abstract}


%%%%%%%%%%%%%%%%%%%%%%%%%%%%%%%%%%%%%%%%%%%%%%%%%%
%%%%%%%%%%%%%%%%% BODY OF PAPER %%%%%%%%%%%%%%%%%%

\section{Introduction}
\label{sec:intro}

\defcitealias{Karastergiou2006}{KJ06}
\defcitealias{Ilie2019}{I19}

A full understanding of the pulsar radio emission mechanism requires explanations of the aspects of the emission that are observed to evolve with frequency. Previous investigations have largely had to be performed by different telescopes at discrete narrowband frequencies, and different epochs in most cases. In these, pulse profiles are seen to broaden with decreasing frequency and individual profile components can shift position and have different spectral indices from each other \citep[eg.][]{Cordes1978, Mitra2002}. Polarization evolution is also observed, with orthogonal jumps present in different numbers and positions at different frequencies, and overall depolarization of the pulse profile tending to happen at higher frequencies \citep[eg.][]{Johnston2006a, Johnston2008}. In addition to the effects seen in the temporally stable integrated pulse profile, single pulses also exhibit a range of frequency-dependent phenomena, for example single pulses of PSR~J1136+1551 have components shifting away from each other at lower frequencies \citep{Oswald2019} and nulling is seen not to be uniform across frequency \citep{Bhat2007}. 
Theories of radius-to-frequency mapping \citep{Ruderman1975, Barnard1986, Sieber1997, Dyks2015} and the differential evolution of orthogonal modes with different spectral indices \citep{McKinnon2000, Karastergiou2005, Smits2006} have been proposed to explain these observations, but the majority of studies have been limited in their scope by the observational capacities of the telescopes used. A complete picture of the emission mechanism, the beam geometry, and frequency dependence of the radio emission in general can only be achieved by filling in the gaps between previous narrowband observations. 

In recent years the LOFAR telescope has provided data at a frequency range of 40--190 MHz, very wide in comparison to its central frequency, which can be used to address some of these issues at low radio frequencies. Work by \cite{Noutsos2015} using LOFAR data and \cite{Hassall2011} using simultaneous observations with LOFAR, the Lovell (1400 MHz), and Effelsberg (8000 MHz) Telescopes has addressed profile evolution with frequency. The Ultra-Wideband receiver on the Parkes radio telescope (UWL) \citep{Hobbs2019a} provides for the first time continuous broadband observations across a 700--4000~MHz band. This is a complementary observing range to LOFAR, and one which covers the 1400~MHz band at which the majority of narrowband pulsar observations have been made in the past. 

Observational data contain signatures of propagation through the interstellar medium in the form of dispersion, temporal scattering and Faraday rotation, in addition to the intrinsic phenomena discussed.
The dispersion measure (DM) of a newly-discovered pulsar is initially identified through alignment of the intensity profile, by finding the DM which maximizes the signal-to-noise ratio (SNR) of the total power profile summed across frequency. In the context of pulsar timing, a higher precision and often time-variable DM can be computed using frequency resolved templates \citep[e.g.][]{Liu2014,Pennucci2018}. Neither technique for establishing the DM addresses the question of how to relate the rotational phase of observed pulse profiles at different frequencies; even frequency resolved profiles are phase aligned in an arbitrary way.
The DM used to align previous multi-frequency data has also been defined in a variety of ways, e.g. alignment of core components or centroids of symmetrical pairs \citep{Hankins2008}.
The rotation measure (RM) is calculated from dedispersed polarization data: methods for doing so include fitting a wavelength-squared relationship to the polarization position angle (PA) \citep[e.g.][]{Rand1994,Noutsos2009}, and finding the peak in the Fourier transform of the polarization intensity vector with respect to wavelength squared, known as RM synthesis \citep{Brentjens2005, Sobey2019}. Crucially, the best fit RM for a given pulsar will depend on how the data have been dedispersed. It is therefore important to fit for both of these parameters together and to ensure that the DM applied aligns the polarization appropriately. The RM calculation method used by \cite{Ilie2019}(\citetalias{Ilie2019}) involved selecting the DM for which the variations of the phase-resolved RM across the profile were minimized. Any remaining RM variations across the profile they then associated with either interstellar scattering or magnetospheric effects. 

Our goal is to observe how pulsar emission evolves across frequency, and so to relate our understanding of this evolution to the three-dimensional structure of the pulsar beam. 
For this we need our observations to be aligned across frequency in a meaningful way, and corrected of interstellar propagation effects. Within this context we identify three key questions that we address in this paper.
\begin{enumerate}
	\item Does the shape of the pulsar profile, including the magnitude and direction of the linear polarization, evolve continuously with frequency, or are there discontinuous changes?
	\item Is it possible to align the PA profile of a pulsar across frequency when constrained to do so using only the RM and the DM?
	\item If so, does this alignment also result in alignment of the total power? Can misalignment be explained in terms of theories of the pulsar beam structure and emission height?
\end{enumerate}

There is a broad consensus that the shape of the PA profile of a pulsar results from its underlying geometry. Under this assumption, it should remain unchanged with observing frequency, making it possible to compare, and so align, PA profiles at different frequencies. Deviations from this behaviour are known to exist in the form of orthogonal modes of polarization, and other features reported in the literature \citep[e.g.][]{Johnston2017c}. These often stem from complex, frequency-dependent  distributions of single pulse polarization states \citep[e.g.][]{simobs1,simobs2,simobs3}. 
Nevertheless, across the observed population the PA appears constant with frequency. \cite{Karastergiou2006} (\citetalias{Karastergiou2006}) showed the difference between the PA profiles of 17 pulsars at 1.4 and 3.1 GHz is tightly distributed around zero. Similarly, fig. 1 in \cite{Mitra2004a} shows very consistent PA profiles at frequencies between 0.4 and 1.6~GHz for the pulsars studied in that paper. This is an observational result that we seek to exploit in our analysis.

Pulsar theory predicts a lag in the arrival of the PA with respect to the intensity profile due to aberration and retardation (A/R). Emission produced at a height $r$ above the pulsar surface will have a relative time delay of 
\begin{equation}
\delta t = 4\frac{r}{c},
\label{eq:1}
\end{equation}
as first described by \cite{Blaskiewicz1991}, where $c$ is the speed of light.
If the emission height of the radio beam is frequency-dependent (commonly referred to as radius-to-frequency mapping, RFM hereafter), this time delay will also vary with frequency. In the literature, A/R effects are measured by identifying the phase of the centre of the pulse profile, the inflection point of the PA, and the lag between them. The complexity of pulsar profiles, including the evidence that pulsar beams seem rarely to be completely filled, means that it is often difficult to choose the appropriate pulse phase on the intensity profile to perform this comparison. 
The PA inflection point is also difficult to measure in pulsars with complex PA profiles. 
Despite these caveats, estimates using this method suggest pulsar emission originates from heights in the range $\sim$10--1000~km \citep[e.g.][]{Blaskiewicz1991, Mitra2004a, Johnston2019a} with heights below 400~km being favoured in the literature \citep{Mitra2002, JK2019}.

Broadband polarimetric observations allow us the opportunity to investigate frequency dependent delays between total power and polarization profiles. To achieve this, we develop a new method for alignment of PA profiles across frequency. Past work, e.g. \cite{Mitra2004a} and \citetalias{Karastergiou2006}, has focused on finding the best alignment for PA profiles and subsequently interpreting the shifts required to do so in terms of theoretical properties of pulsar emission. We take a different approach, restricting shifts of the PA in phase and angle to follow the frequency-squared dependencies of dispersion and Faraday rotation. We then scan over DM and RM simultaneously to find the pair of values that give the best alignment. For this work we require broadband data in order to identify the frequency-squared behaviour of the interstellar propagation effects. This allows us first to discover whether it is possible to find an alignment for a given pulsar when restricted in this way, and secondly gives us the frequency resolution required to study intrinsic properties of the emission, once we have obtained such an alignment. The unprecedented factor of 6 frequency span of observations made with the UWL results in ideal data for this work. 

Pulsar emission is frequently observed to have a circularly polarized component, the strength of which is commonly frequency-dependent, as expected under the assumption that observed circular polarization is a consequence of propagation effects in the pulsar magnetosphere. It follows that, in cases where the circular polarization is seen to vary considerably with frequency, the PA profile will also display frequency evolution, challenging the methodology we present here. In general however, the magnitude of such changes in circular polarization is low, suggesting that our method could be applicable to large numbers of sources. Furthermore, even in cases of variations of greater magnitude, the alignment process described here provides a starting point from which the frequency-dependent deviations due to more complex polarization effects can be separated from the effects of interstellar propagation and investigated more carefully.

We present in this paper the alignment and visualization of two example pulsars observed with the UWL: PSR~J1359--6038 and PSR~J1056--6258. \citetalias{Karastergiou2006} showed, for both of these sources, that aligning two PA profiles at two spot frequencies, 1.4 and 3.1~GHz, resulted in unexplained offsets of the intensity profiles. Such misalignment might have been intrinsic to the pulsar emission processes, or may have been due to over-emphasis in the alignment of small differences in the PA profile shape local to each of the two frequencies used. The new observations and analysis presented here can provide an explanation.

The paper is structured as follows. A description of the observing strategy and data processing in section \ref{sec:obs} is followed by an explanation of the method developed for aligning broadband PA profiles using RM and DM corrections. The results for both simulations and the test case pulsars are presented and discussed in section \ref{sec:RMDM}. In section \ref{sec:UWL} we present the UWL data for PSRs J1359--6038 and J1056--6258, having applied our calculated RM and DM. Conclusions are summarized in section \ref{sec:conc}.

\section{Observations and data processing}
\label{sec:obs}
\begin{figure*}
    \includegraphics[width=\textwidth]{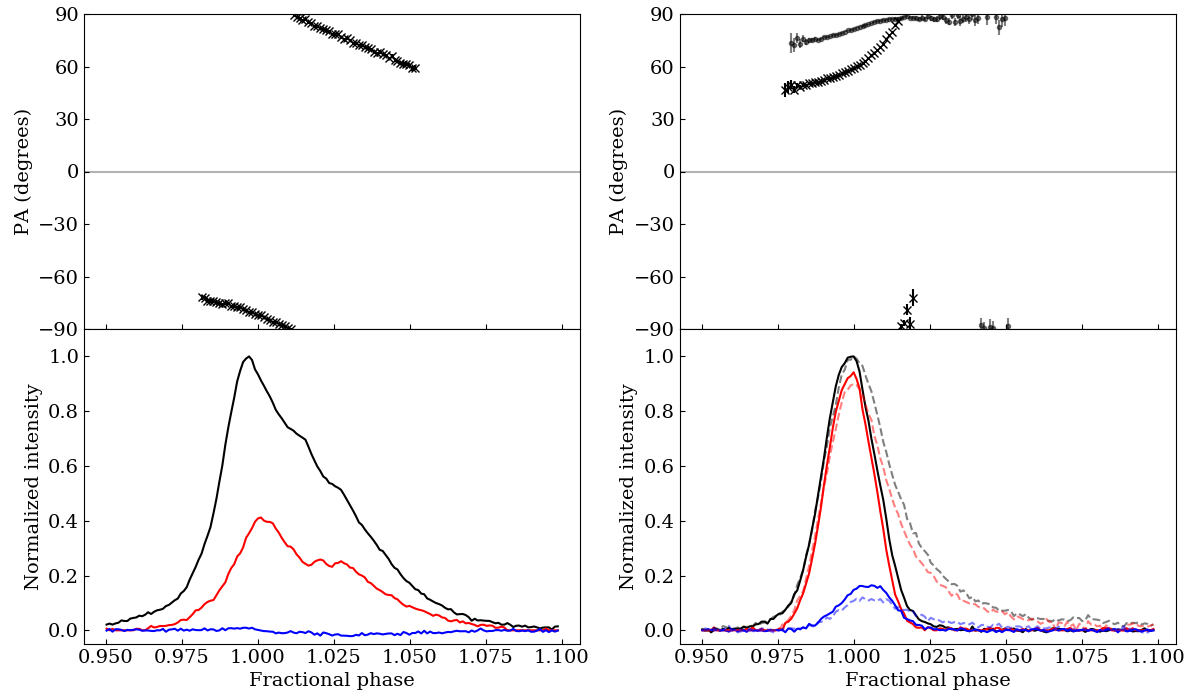}
    \caption{Integrated pulse profiles for PSR~J1056--6258 at 1410~MHz (left) and for PSR~J1359--6038 at 1410~MHz and 830~MHz (right). The top half of each graph shows the absolute PA profile, showing only those values for which $L_{\rm SNR} > 20$ (left) and $L_{\rm SNR} > 5$ (right), with points marked by black crosses with errorbars. The bottom half shows the total intensity (black), linear polarization (red) and circular polarization (blue). For PSR~J1359--6038 the profile at 830~MHz is overlaid, with faint dashed lines for the profile and faint dots with error bars for the PA, to show the effect of temporal scattering on the profile shape at lower frequencies. The RM (33~rad~m$^{-2}$) and DM (293.736~cm$^{-3}$pc) used to align these two frequencies for the figures were taken from the pulsar catalogue \citep{Manchester2005}. It is immediately clear that the catalogue RM is incorrect, since the two PA profiles do not overlap on the unscattered left-hand side of the profile.
    }
    \label{fig:1}
\end{figure*}

A programme of observations of non-recycled pulsars has been running on the Parkes radio telescope since 2007 under project P574. Prior to late 2018, the prime receiver for the project was the Parkes Multibeam, operating at a frequency of 1369~MHz and a bandwidth of 256~MHz \citep[see for example][]{Johnston2017c}. As of November 2018, the programme uses the UWL receiver in conjunction with the Medusa backend which extends the frequency range to run between 700 and 4000~MHz \citep{Hobbs2019a}. A total of 250 pulsars are observed on a monthly basis. The data are coherently dedispersed at the published pulsar DM, folded at the topocentric spin-period of the pulsar, divided into 1024 phase bins and integrated for 15~s for each of 3328 frequency channels of 1~MHz bandwidth, and written to disk. Observations of a pulsed signal, injected into the feed, are also recorded in order to determine the relative gains and phases between the two probes of the receiver. Observations of Hydra~A are made to determine the absolute flux scale. Data reduction involves flagging of edge-channels and data corrupted by interference and application of the gain, phase and flux calibration parameters. The data are then summed over time and aggregated in frequency to provide 8 calibrated profiles across the band in each of the Stokes parameters. The centre frequencies and bandwidths of these 8 frequency-aggregated profiles are given in table \ref{tab:1}. Profiles for the two pulsars analysed here, PSRs~J1359--6038 and J1056--6258, are shown in Fig. \ref{fig:1}.

\begin{table}
\caption{Centre frequencies and bandwidths of the 8 frequency channels used for aligning the observational data across frequency.}
\label{tab:1}
\begin{tabular}{ccc}
 \hline
\textbf{Channel} & \textbf{Centre frequency} & \textbf{Bandwidth} \\
\textbf{index}& \textbf{(MHz)} & \textbf{(MHz)} \\
 \hline
 1 & 830 & 256\\
 2 & 1080 & 256\\
 3 & 1410 & 384\\
 4 & 1790 & 384\\
 5 & 2110 & 256\\
 6 & 2750 & 512\\
 7 & 3260 & 512\\
 8 & 3780 & 512\\
 \hline
\end{tabular}
\end{table}

\section{Process of PA alignment}
\label{sec:method}

We calculate the $\chi ^{2}$ value associated with the alignment of PA profiles across the 8 frequency channels, on the application of $\Delta$RM and $\Delta$DM, corrections to the catalogue values of RM and DM. This is done as follows:

\begin{enumerate}
    \item de-Faraday and dedisperse Stokes parameters Q and U with a pair of $\Delta$RM and $\Delta$DM values.
    \item Calculate the PA as $0.5\arctan(U/Q)$ and remove those channels where the presence of strong scattering or orthogonal mode jumps alters the PA profile sufficiently that it is not comparable in shape to the profile at the other frequencies.
    \item Calculate $\chi ^{2}$, using only those bins where the SNR of the linear polarization $L_{\rm SNR}$ is greater than a defined limit for all frequency channels.
\end{enumerate}

The linear polarization $L = \sqrt{Q^{2} + U^{2}}$ is bias-corrected as described in \cite{Everett2002} before $L_{\rm SNR}$ is calculated. 
We calculate $\chi ^{2}$ in terms of $d\rm{PA}$, the angular distance of each PA profile from the PA profile at the lowest frequency, which is minimized when $\Delta$RM and $\Delta$DM align the data correctly. The choice of calculating the alignment with respect to the lowest frequency is due to the spectral index of pulsar intensity, meaning that the profile at the lowest frequency will have the highest SNR and therefore the largest number of contributing data points with a SNR greater than the cut-off.
We define $d\rm{PA}$ allowing for cyclic boundary conditions, so that for the $i$th frequency channel (where $i=0$ is the lowest frequency) and $j$th phase bin it is given by
\begin{equation}
    d\rm{PA}_{\rm ij} = \left(\left(\abs{\rm{PA}_{\rm ij} - \rm{PA}_{\rm 0j}} + \frac{\uppi}{2}\right)\mathbin{\%}\uppi - \frac{\uppi}{2}\right)\times sign(\rm{PA}_{\rm ij} - \rm{PA}_{\rm 0j}).
\end{equation}
The error associated with a value of $d$PA depends on the errors of the two contributing PA values as
\begin{equation}
    \sigma_{\rm ij} = \sqrt{\sigma_{\rm{PA}_{\rm ij}}^{2} + \sigma_{\rm{PA}_{\rm 0j}}^{2}},
\end{equation}
where $\sigma_{PA}$ is calculated as in \cite{Everett2002}.
Summing over frequency channels and phase bins then gives a normalized $\chi ^{2}$ of
\begin{equation}
    \chi^{2} = \frac{1}{N_{\rm chan}N_{\rm bin}}\sum_{\rm i}^{N_{\rm chan}}\sum_{\rm j}^{N_{\rm bin}}\frac{dPA_{\rm ij}}{2\sigma_{\rm ij}^{2}},
\end{equation}
so that the likelihood function for the PA alignment is $\exp(-\chi^{2})$.

We sample the log-likelihood function over the parameter space of $\Delta$RM and $\Delta$DM using the MCMC algorithm {\sc emcee} \citep{Foreman-Mackey2013}.

\section{Rotation and Dispersion Measures that align the PA}
\label{sec:RMDM}

\subsection{Simulations}

In order to test the precision and accuracy of our alignment method, we developed three simulations of pulsar data with increasing levels of complexity. The first and simplest simulated pulsar has a symmetrical intensity profile and strongly curved PA profile, with geometry $\alpha = 100\degr$ and $\beta = -2\degr$, where $\alpha$ is the angle between the pulsar magnetic and rotation axes and $\beta$ is the impact parameter of our line of sight. We designed our second simulation to be similar to PSR~J1056--6258, to investigate how well the alignment technique could handle a flatter PA ($\alpha = 156\degr$, $\beta = -10.5\degr$) and an asymmetric profile. The third simulation includes the observational effect of frequency-dependent emission heights and is otherwise identical to the second.

For each simulation, the intensity profile is a Gaussian (either symmetrically or asymmetrically placed) with flux spectral index $-1.4$. The linear polarization is an identical Gaussian with peak amplitude half that of the intensity profile at each frequency. We added Gaussian noise with a standard deviation $\sigma_{\rm I}$ given by the corresponding $\sigma$ value for the off-pulse noise at each frequency for PSR~J1056--6258. Hereafter when we refer to the curvature of the PA profile we are giving a qualitative description intended to convey whether the gradient of the PA profile varies considerably with phase (strong curvature) or not (weak or little curvature). We dispersed and Faraday rotated each simulation such that $\Delta$RM = $+2$~rad~m$^{-2}$ and $\Delta$DM = $+1$~cm$^{-3}$pc were required to realign the profiles.

For the third simulation we applied height-dependent shifts to the intensity and PA profiles based on A/R, and we used the model developed in \cite{Oswald2019} to further calculate how the intensity profile evolves with frequency as emitting field lines diverge. We used the height-frequency relationship calculated in that paper for PSR~J1136+1551 of $r(f)\propto f^{-0.39}$, with height $r$ in km and frequency $f$ in GHz. This relation comes from applying the fan beam model to the single-pulse frequency evolution of PSR~J1136+1551 across the frequency range 330--470~MHz, assuming a 1-to-1 relationship between frequency and emission height and the field lines of the emitting region having a footprint parameter ratio, as defined in \cite{Gangadhara2001}, of 0.5. The particular height frequency relationship and footprint parameter ratio are only taken as indicative examples. 

We followed the prescription of \cite{Dyks2008} to implement A/R in our third simulation: the intensity and PA profiles are shifted in phase by equal amounts of $2\frac{r}{R_{\rm LC}}$ earlier and later respectively, where $R_{\rm{LC}}$ is the radius of the light cylinder of the pulsar, given in terms of the pulsar period as $\frac{cP}{2\uppi}$. This is found to be valid for a range of theoretical models provided emission is produced at heights below 0.1~$R_{\rm{LC}}$ \citep{Craig2012}. 
\cite{Dyks2008} also describes a height-dependent PA rotation of $\frac{10}{3}\frac{r}{R_{\rm LC}}\cos{(\alpha)}$ as calculated by \cite{Hibschman2001}, which we include in the simulation. The PA rotation between the top and bottom of the band for the heights mentioned above (assuming a pulsar rotational period of 0.42~s as for PSR~J1056--6258) is $1.8\degr$, which is small compared to the effect of Faraday rotation. 

\begin{figure*}
    \includegraphics[width=\textwidth]{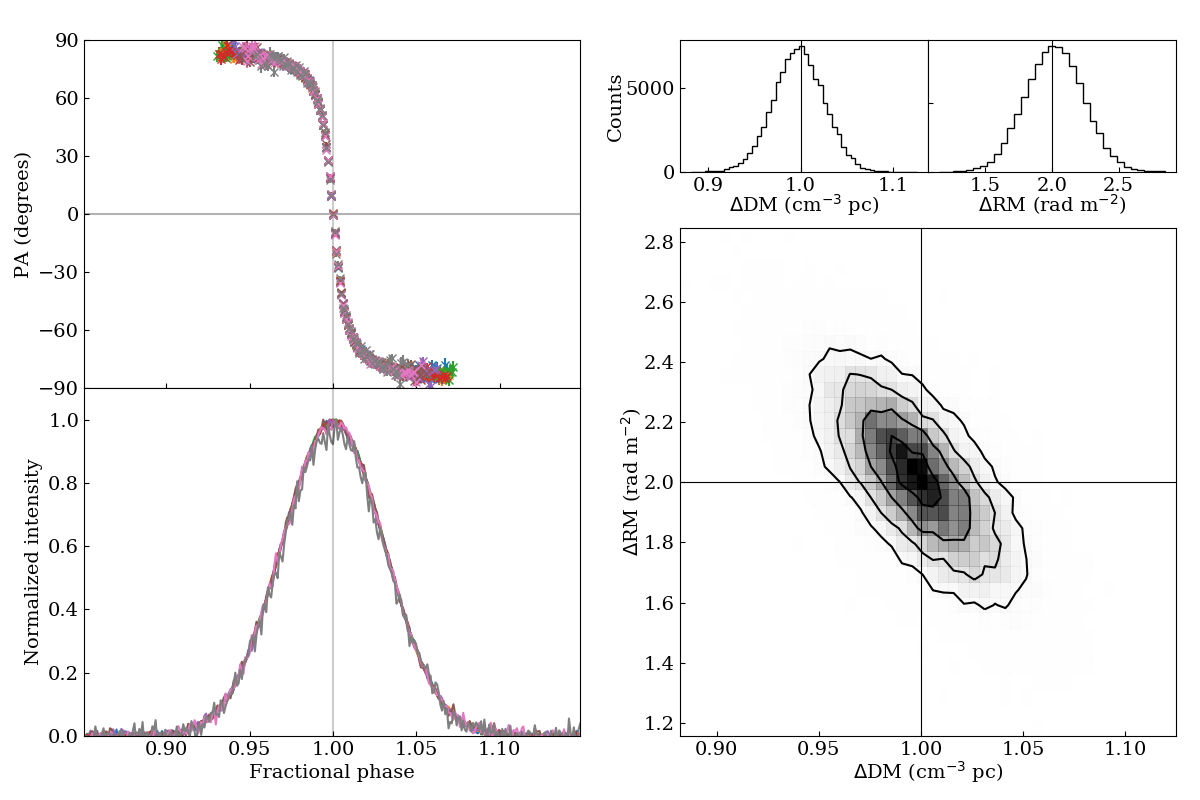}
    \caption{Plots showing the alignment of a simulation of pulsar emission with a PA geometry of $\alpha = 100\degr$, $\beta = -2\degr$. Right: the main plot shows a 2D histogram, showing the alignment probability distribution for pairs of parameters $\Delta$RM and $\Delta$DM. The vertical and horizontal black lines overlaid indicate the true values of these parameters for the simulation: $\Delta$RM = 2~rad~m$^{-2}$ and $\Delta$DM = $1$~cm$^{-3}$pc. Above are the marginalized distributions of $\Delta$DM and $\Delta$RM with the true values marked with vertical black lines. Left: simulated pulsar profiles aligned using the $\Delta$RM and $\Delta$DM values marked on the 2D histogram, with all 8 frequencies shown. The frequencies are overlaid from lowest to highest, so that the lowest frequency of 827~MHz (coloured blue) is plotted first and the highest frequency of 3775~MHz (coloured grey) is plotted last. The upper panel shows the position angle and the lower panel shows the normalized total intensity. }
    \label{fig:2}
\end{figure*}

The left-hand side of Fig. \ref{fig:2} shows the first simulation, with the intensity and PA profiles plotted for all 8 frequency channels. The right-hand side of the figure shows the 2D histogram of PA alignment probability over the parameter space of $\Delta$RM and $\Delta$DM, with the marginalized histograms shown above. It is tightly constrained and the peak lies at the correct values.

\begin{figure*}
    \includegraphics[width=\textwidth]{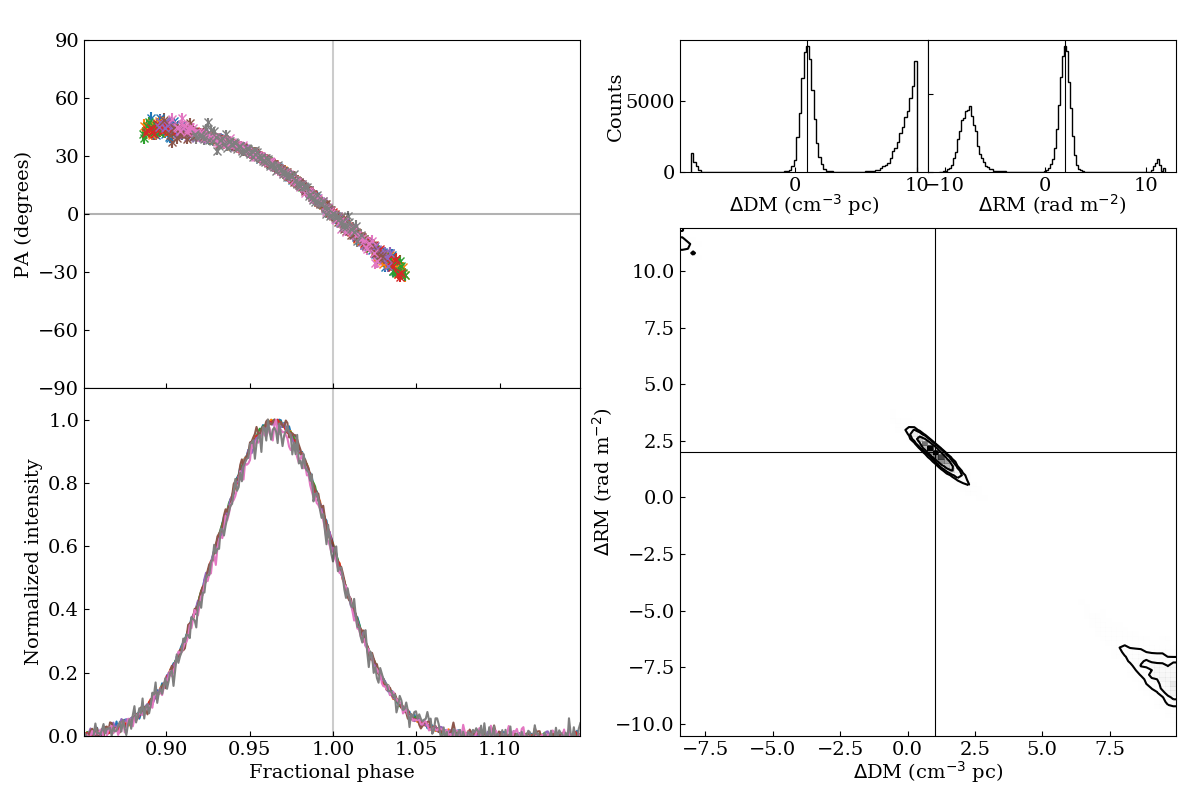}
    \caption{Simulation of pulsar emission designed to be similar to the observation of PSR~J1056--6258. See Fig. \ref{fig:2} for details of the figure contents.}
    \label{fig:3}
\end{figure*}

The 2D histogram for the second simulation (Fig. \ref{fig:3}) is again peaked at the true alignment parameter values; the correctly alignment is successfully recovered. This histogram is wider in comparison to the last: the weaker curvature of the PA means that the PA profiles can be shifted diagonally with respect to each other and still remain well aligned, meaning that the error bounds on the alignment probability increase. Furthermore, there are non-negligible regions of alignment probability at large shifts of $\Delta$RM and $\Delta$DM, forming notable side-lobes in the marginalized posterior distributions. These regions correspond to the PA profiles being shifted so far relative to each other that only a few data points are overlapping. In addition, these data points, being near the edges of the intensity profile, have lower SNR of the linear polarization and hence larger error bars, making alignment still easier. It is clear that, although it is possible to obtain a good alignment probability when including so few data points, such an alignment does not correspond to a real overlap of the PA curve. In the data analysis we must therefore focus on the central component of the marginalized posterior distributions to obtain uncertainties.

\begin{figure*}
    \includegraphics[width=\textwidth]{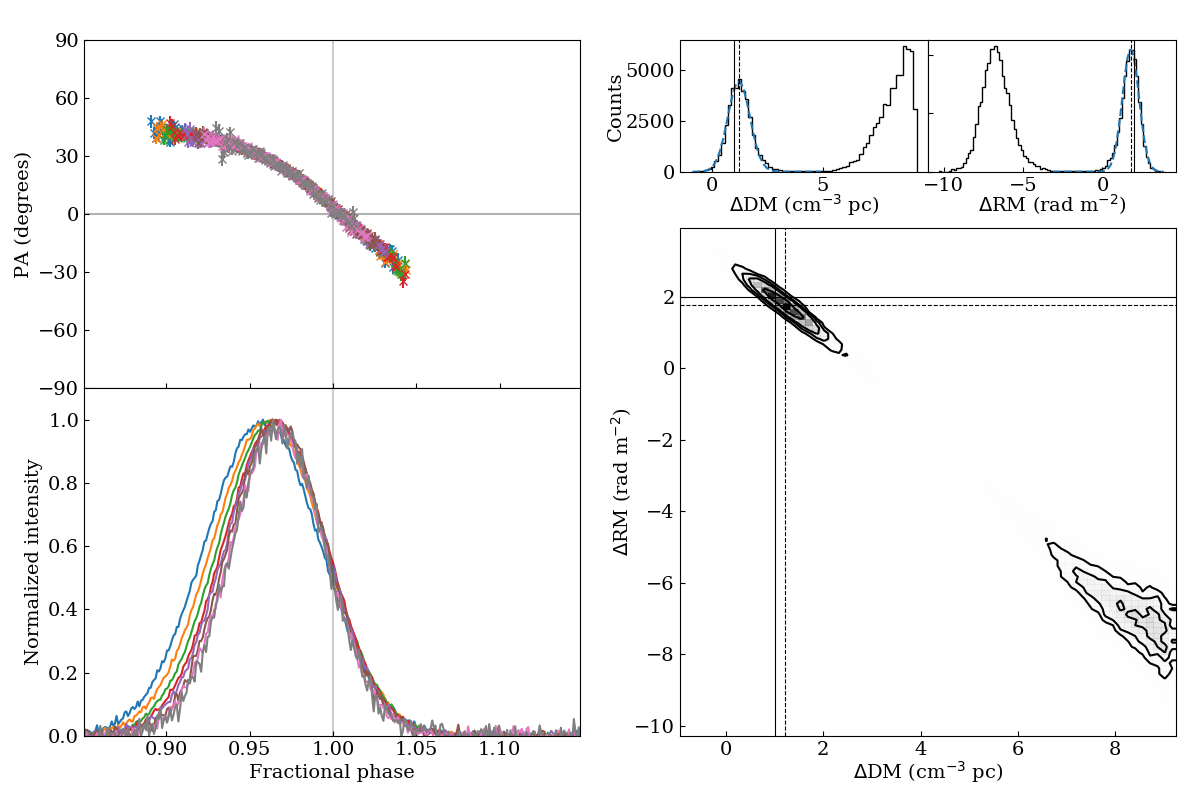}
    \caption{Simulation as in Fig. \ref{fig:3} but now including the effects of varying emission height with frequency. The alignment used to plot the profiles on the left is the best fit indicated by dotted horizontal and vertical lines on the 2D histogram, found by fitting Gaussians to the marginalized distributions (blue dashed lines). The true values of $\Delta$RM = 2~rad~m$^{-2}$ and $\Delta$DM = $1$~cm$^{-3}$pc are indicated on the histograms with solid lines.}
    \label{fig:4}
\end{figure*}

Fig. \ref{fig:4} shows the alignment of the height-varying model based on the PA. We fit Gaussians to the marginalized $\Delta$DM and $\Delta$RM distributions (marked in blue) to identify the best fit and its error. As for the previous simulation we neglect the probability region corresponding to a large $\Delta$DM and clear misalignment. We further note that only the positive $\Delta$DM side-lobe is seen in the marginalized posterior distribution, as was almost the case in Fig. \ref{fig:3}. In this simulation, the peaks of $\Delta$DM and $\Delta$RM do not correspond to the true values for these parameters. $\Delta$DM stems from A/R shifting the PA with respect to the total power profile in pulse phase, whereas $\Delta$RM additionally stems from
the simulated PAs including both Faraday rotation and rotation due to different emission heights. Shifting the profiles based on the best recovered $\Delta$DM and $\Delta$RM results in aligned PA profiles but a frequency-dependent offset in the total power. We would expect to see a similar effect in real data as a consequence of A/R.

\subsection{Data}

Mindful of the effects of PA curvature and A/R on the PA alignment probability as demonstrated through the simulations, we now fit for the RM and DM that best align the PA for pulsars PSR~J1359--6038 and PSR~J1056--6258. In Table \ref{tab:2} we give the catalogue RM and DM values and the corrections to them required to align the PA, along with the $\Delta$RM and inferred $\Delta$DM values from \citetalias{Karastergiou2006} and \citetalias{Ilie2019}, which are presented for comparison. 

\begin{table*}
    \caption{The DM and RM for the pulsars studied in this paper, given as the catalogue values and the difference to these from this work and from that of both \citetalias{Karastergiou2006} and \citetalias{Ilie2019}. We calculated the values of $\Delta$DM for \citetalias{Karastergiou2006} by estimating the DM required to align the intensity profiles at the frequencies they used and then converting the time delay given in the paper to its associated $\Delta$DM, given the observational frequencies. We converted the $\Delta$DM corrections from \citetalias{Ilie2019} to be defined with respect to the catalogue values and used their average of their phase-resolved RM calculations to obtain $\Delta$RM. The values of $\dot{E}$ are from \protect\cite{Manchester1978}. The catalogue values for RM and DM have the following references. PSR~J1359--6038 DM: \protect\cite{Petroff2013}; RM: \protect\cite{Han1999}. PSR~J1056--6258 DM: \protect\cite{Alessandro1993}; RM: \protect\cite{Costa1991}.}
\label{tab:2}
\renewcommand{\arraystretch}{1.2}
\setlength{\tabcolsep}{3pt}
\begin{tabular}{cccccccccc}
 \hline
 \textbf{Name}   & \textbf{$\dot{E}$} & \textbf{DM$_{cat}$} & \multicolumn{3}{c}{\textbf{$\Delta$DM} (cm$^{-3}$pc)} & \textbf{RM$_{cat}$} & \multicolumn{3}{c}{\textbf{$\Delta$RM} (rad~m$^{-2}$)} \\
 & (ergs s$^{-1}$) & (cm$^{-3}$pc) & \citetalias{Karastergiou2006} (est.) & \citetalias{Ilie2019} & This work & (rad~m$^{-2}$) & \citetalias{Karastergiou2006} & \citetalias{Ilie2019} & This work \\
 \hline
 J1359--6038 & $1.2\times10^{35}$ & $293.736\pm0.003$ & $+0.4$ or $0$ & $0.498\pm0.004$& $+0.2\pm0.2$ & $33\pm5$ & $+6$ or $+4\pm0.5$ & $5.5\pm0.1$ & $+6^{+0}_{-2}$ \\
 J1056--6258 & $1.9\times10^{33}$ & $320.3\pm0.6$ & $+6.6$ or $+0.5$ & $0.79\pm0.01$ & $+1.5\pm0.8$ & $4\pm2$ & $-5$ or $+2\pm0.5$ & $2.5\pm0.1$ & $+2\pm1$ \\
 \hline
\end{tabular}
\renewcommand{\arraystretch}{}
\end{table*}

\subsubsection{PSR~J1359--6038}

PSR~J1359--6038 is affected by strong temporal scattering which shifts emitted power to later phases, resulting in a scattering tail for the intensity profile and the later part of the PA profile being flattened, as described by \cite{Karastergiou2009a}. However, as scattering is strongly frequency-dependent, the deformation of the PA is only important at the lowest observed frequencies. We therefore exclude channels 1--3 (see Table \ref{tab:1}) from the fitting process and only plot them subsequently, at which point we can use them to constrain the RM correction. We perform the fit using the PA values with $L_{\rm SNR} > 5$.

\begin{figure*}
    \includegraphics[width=\textwidth]{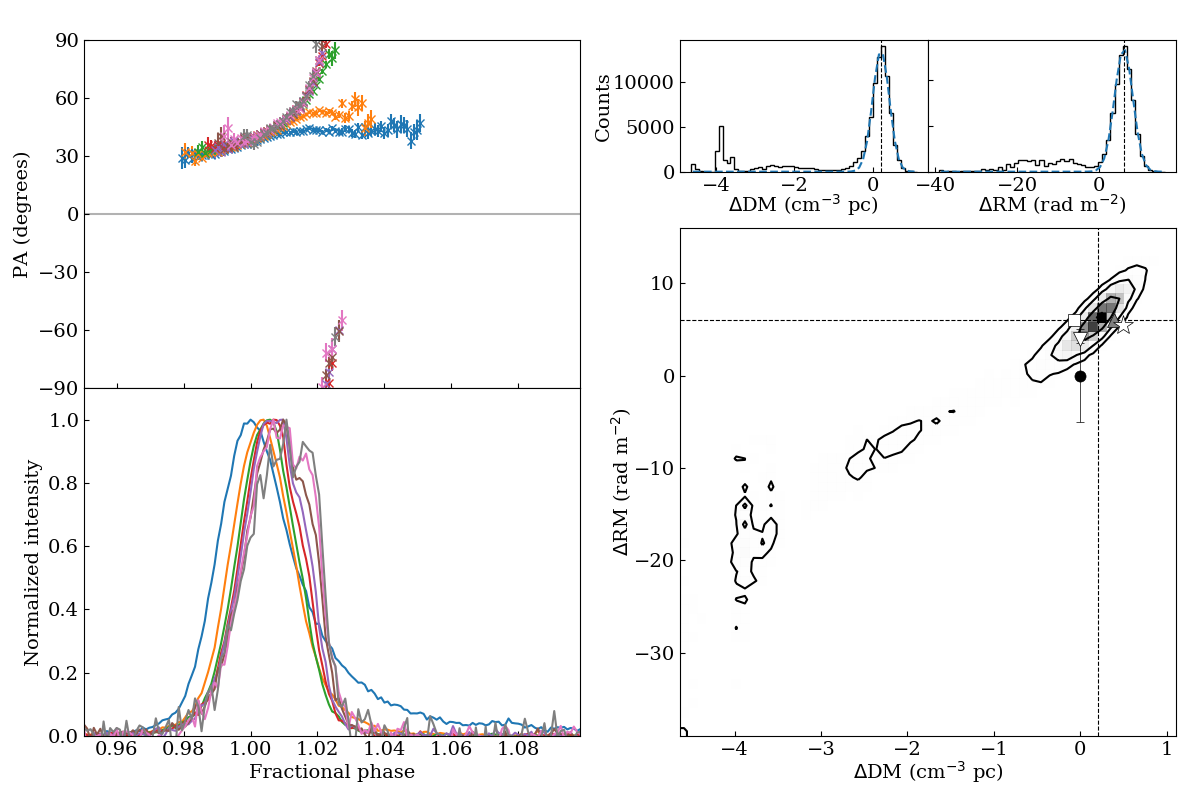}
    \caption{Plots as described in Fig. \ref{fig:2}, now showing PSR~J1359--6038. The alignment on the left is $\Delta$RM = $+6$~rad~m$^{-2}$ and $\Delta$DM = $+0.2$~cm$^{-3}$pc, identified by Gaussian fits to the marginal $\Delta$DM and $\Delta$RM histograms (blue dashed line), and marked with dashed lines on the 2D histogram. Overlaid on the 2D histogram are the $\Delta$RM and $\Delta$DM values from the catalogue (black circle) and those computed by \citetalias{Karastergiou2006} aligning by intensity (white inverted triangle) and PA (grey triangle). The corrections calculated by \citetalias{Ilie2019} are marked with a white star. An additional marker shows the value of $\Delta$DM = $-$0.075~cm$^{-3}$pc calculated from a scattering fit to the intensity profiles (white square).}
    \label{fig:5}
\end{figure*}

Fig. \ref{fig:5} shows the alignment results for PSR~J1359--6038. Similar to Fig. \ref{fig:3}, there is some probability associated with large DM shifts which can be neglected. As described for the height-varying simulation, we fit Gaussians to the marginal $\Delta$DM and $\Delta$RM distributions (marked with blue dashed lines) to obtain the best fit $\Delta$DM = $+0.2\pm 0.2$~cm$^{-3}$pc and $\Delta$RM = $+6\pm2$~rad~m$^{-2}$. The best fit is marked on the 2D histogram with vertical and horizontal dashed lines. The catalogue values and those calculated by \citetalias{Karastergiou2006} and \citetalias{Ilie2019} are indicated with markers. The catalogue values lie outside the contoured region whilst the two fits by \citetalias{Karastergiou2006} (with respect to the intensity and PA profiles) both lie close to the contour peak, as does the result of \citetalias{Ilie2019}. The plot of the PA profiles on the left-hand side of the figure shows that alignment has been successful even for the three lowest frequencies, which were not included in the fitting. Since the lower frequencies are the most sensitive to changes in RM, adding them in provides a tighter constraint. Consequently we quote $\Delta$RM = $+6^{+0}_{-2}$~rad~m$^{-2}$ in Table \ref{tab:2}. 

In this work we find the DM that best aligns the PA across frequency, whereas published DMs typically align the total power profile. For PSR~J1359--6038 these values differ, however there can be only one true value for the interstellar DM, defined as the line integral of the electron column density along the line of sight. This measurement difference of $\Delta$DM must therefore result from an effect unrelated to interstellar dispersion. In the case of our height-varying simulation the frequency-dependent PA lag, caused by A/R in the case of RFM, was absorbed into the best fit DM. This resulted in visible misalignment of the intensity profile across frequency. Under the assumption that A/R is present in the data, and guided by this simulation result, we can associate our measured $\Delta$DM for PSR~J1359--6038 with a difference in emission height across the frequency band.

Neglecting any intensity profile evolution due to field line curvature, we convert $\Delta$DM into a frequency-dependent time lag. The difference between the time lags at the top and bottom of the band we can then use to calculate an emission height range across the band, as in equation \ref{eq:1}. The purpose of doing so is not to measure the absolute emission heights for this pulsar, but to investigate whether the misalignment we see is of a magnitude that is plausibly attributable to the effects of A/R, thus testing the aforementioned assumption. The calculation gives $\Delta r = 90\pm 90$~km over the observing band. At the lowest frequencies, scattering delays the observed profiles by some further $\Delta t$. Correcting the total delay then results in an estimated indicative height range of $\Delta r = 120\pm 120$~km. We can compare this with the corresponding value from the height-frequency relation of $r(f)\propto f^{-0.39}$ that is given in \cite{Oswald2019}. For our observing band this gives $\Delta r = 200$~km, which is comparable in magnitude. 

\citetalias{Ilie2019} proposed that the variation in phase-resolved RM they detected could be explained partly by scattering, but that the lack of corresponding variations in circular polarization may suggest a further magnetospheric origin for the RM variations across the profile. Although we do not investigate circular polarization, our results are in agreement with \citetalias{Ilie2019} on the effect of scattering on the phase-resolved RM. Our technique provides a complementary method to that of \citetalias{Ilie2019} to investigate frequency-dependent effects on polarization beyond dispersion and Faraday rotation.

\subsubsection{PSR~J1056--6258}

\begin{figure*}
    \includegraphics[width=\textwidth]{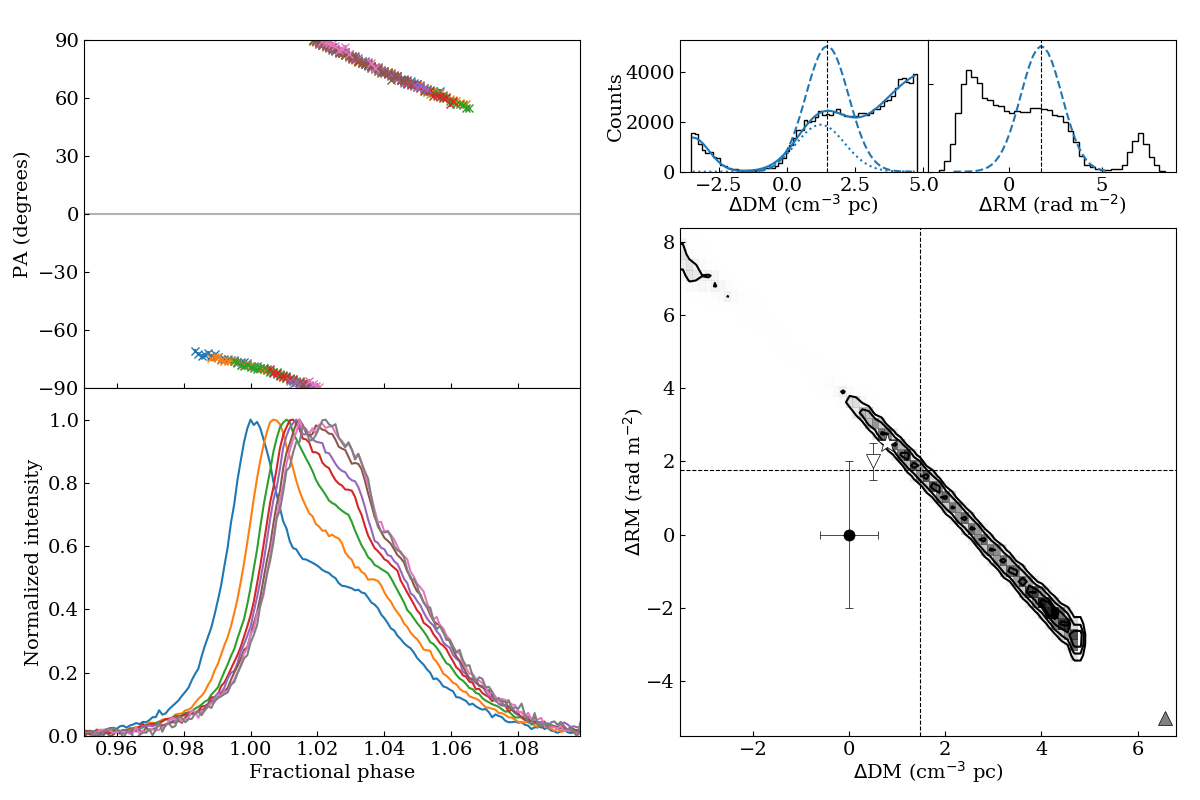}
    \caption{As for Fig. \ref{fig:2}, now showing PSR~J1056--6258. The best fit alignment (shown on the left) is that marked with dashed horizontal and vertical lines on the 2D histogram: $\Delta$RM = $+2$~rad~m$^{-2}$ and $\Delta$DM = $+1.5$~cm$^{-3}$pc. These numbers were obtained by fitting three Gaussians to the marginalized $\Delta$DM distribution (solid blue line) and using the central Gaussian as a prior for a second run of the fitting algorithm. This produced a Gaussian distributed output in $\Delta$RM and $\Delta$DM (dashed blue lines), from which the best fit parameters are obtained. The symbols overlaid on the 2D histogram are as described in Fig. \ref{fig:5}.}
    \label{fig:6}
\end{figure*}

Although unaffected by scattering, at the highest 2 frequency channels the PA profile of PSR~J1056--6258 deviates from the rest with the presence of a pair of orthogonal jumps. We therefore perform the alignment fit for PSR~J1056--6258 using only channels 1--6, with $L_{\rm SNR} > 20$. We see the same edge effects as for the simulation in Fig. \ref{fig:3}, though the lack of curvature of the PA means that they are more dominant. We therefore fit three Gaussians to the marginal $\Delta$DM distribution and use the central Gaussian as our prior on $\Delta$DM. Running the fit a second time with the $\Delta$DM prior gives outputs for $\Delta$DM and $\Delta$RM that follow Gaussian distributions, which we mark on the marginalized distributions in Fig. \ref{fig:6} with dashed blue lines. From these we identify the best $\Delta$DM and $\Delta$RM values and their errors. This gives $\Delta$DM = $+1.5 \pm 0.8$~cm$^{-3}$pc and $\Delta$RM = $+2\pm1$~rad~m$^{-2}$. The catalogue values for RM and DM each individually could lie within the fit region, however, together they lie far outside the contours. This suggests that a different DM may have been used in the calculation of the catalogue RM. The best PA fit of \citetalias{Karastergiou2006} lies along the direction of the contoured region but at too large a shift. It therefore corresponds to the edge effect situation, something that is revealed due to the increased frequency range and resolution of these new data. Their other best fit, based on aligning the intensity profiles and calculating the corresponding RM, sits close to the contoured region. The extension to lower frequencies in these new data provides higher fidelity for the measured RM given the DM, which is likely to explain this discrepancy. The result of \citetalias{Ilie2019} lies in the contoured region close to our best fit parameters.

As for PSR~J1359--6038, the best alignment of the PA for PSR~J1056--6258 does not correspond to a perfectly aligned total intensity. Once again we can compare the DM value that best aligns the total intensity with that which aligns the PA, and associate the difference between these values with an indicative emission height range for the frequency band under the assumption of A/R. The catalogue DM does not appear to align the total intensity well, but we estimate that correcting this value by approximately $+0.5$~cm$^{-3}$pc visually aligns the intensity profile across frequency. The difference between the two DMs aligning intensity and PA respectively is then $\Delta$DM = $+1$~cm$^{-3}$pc. Performing the same calculation as for PSR~J1359--6038, this corresponds to an emission height range of $\Delta r = 430\pm350$~km, which is larger but still comparable in magnitude to the simulated value of $\Delta r = 200$~km.

\section{Presenting broadband polarimetric data}
\label{sec:UWL}

\begin{figure*}
    \includegraphics[width=\textwidth]{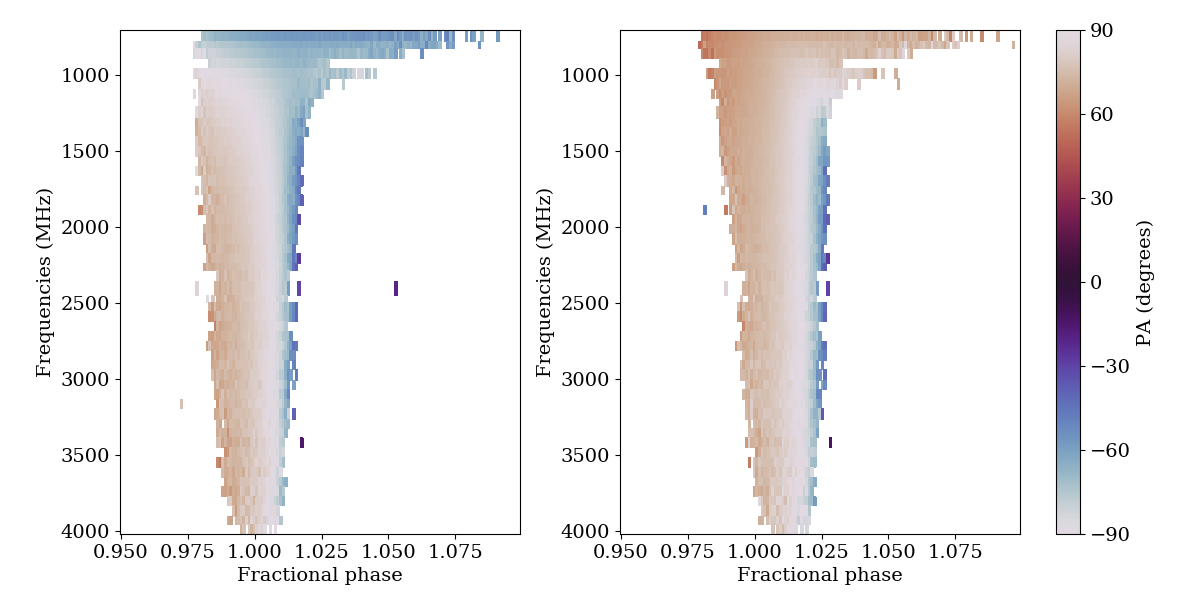}
    \caption{Graph showing how the PA profile varies with frequency for PSR~J1359--6258. Left: aligned using the catalogue RM and DM. Right: aligned using the DM and RM correction picked out in Fig. \ref{fig:5}. The PA is only shown where the SNR of the linear polarization exceeds 4.}
    \label{fig:7}
\end{figure*}

\begin{figure*}
    \includegraphics[width=\textwidth]{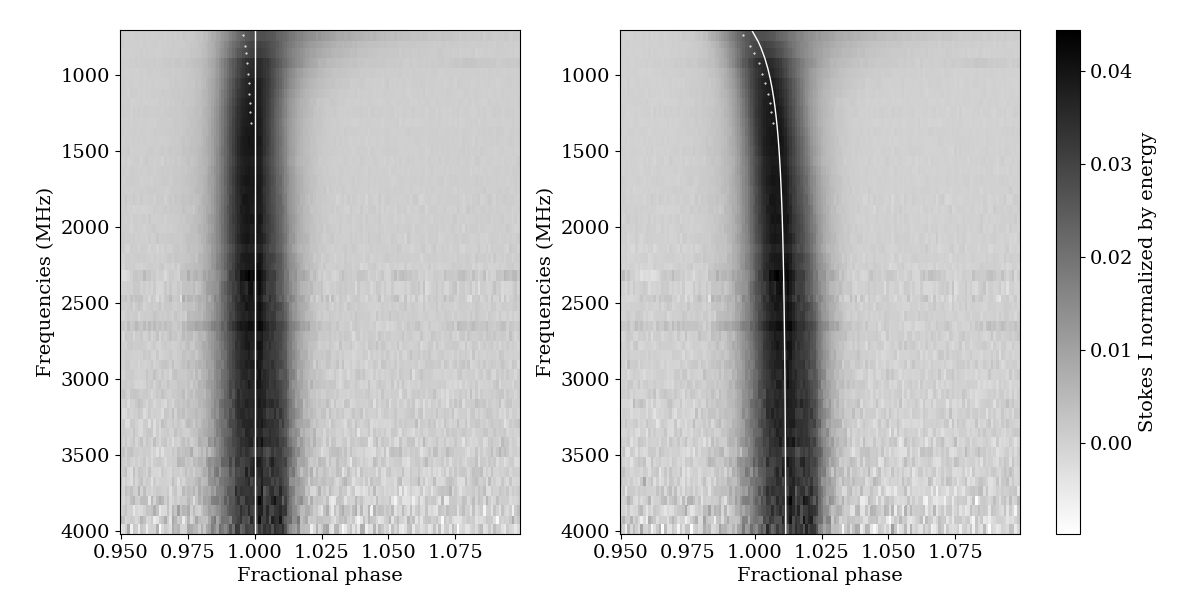}
    \caption{Graph showing how the intensity profile varies with frequency for PSR~J1359--6038, with DM and RM applied as in Fig. \ref{fig:7}. The intensity is normalized by dividing by the summed flux of the on-pulse region. The white line indicates the profile peak and the dots indicated the peak of the unscattered profile for the 10 lowest frequency channels, estimated by fitting to the profile a Gaussian convolved with an exponential scattering function.}
    \label{fig:8}
\end{figure*}

\begin{figure*}
    \includegraphics[width=\textwidth]{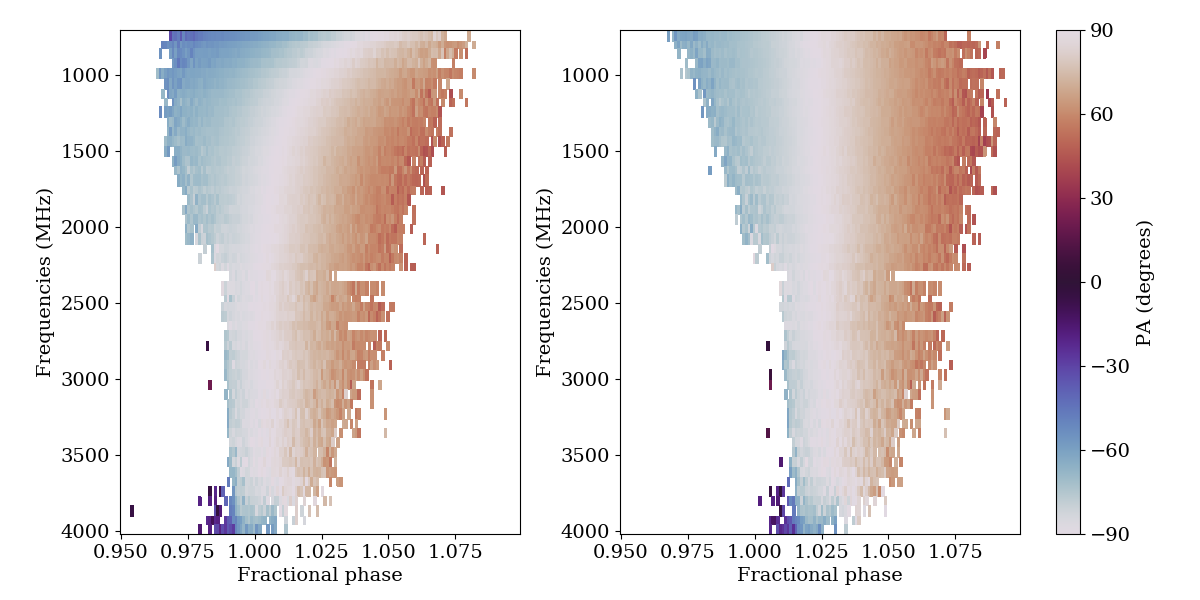}
    \caption{The PA profile for PSR~J1056--6258, with the RM and DM values from the catalogue (left) and this work (right) respectively.}
    \label{fig:9}
\end{figure*}

\begin{figure*}
    \includegraphics[width=\textwidth]{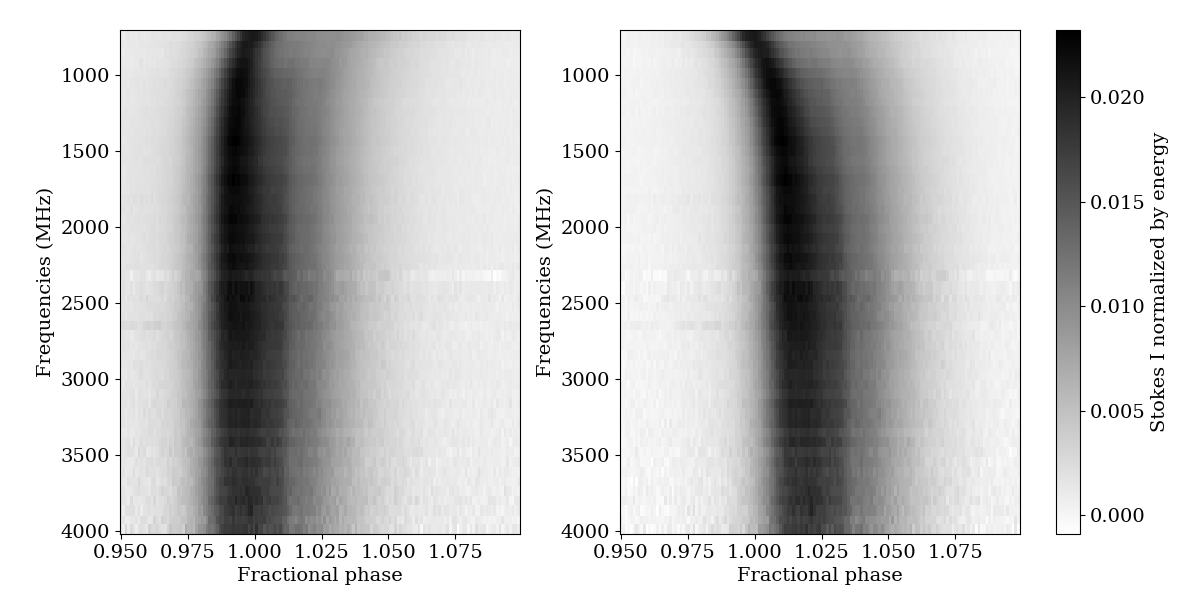}
    \caption{The intensity of PSR~J1056--6258, normalized by dividing by the on-pulse flux. DM and RM values applied are as in Fig. \ref{fig:9}.}
    \label{fig:10}
\end{figure*}

Having obtained RM and DM for each of our pulsars, we reprocess the data using these new parameters. We then aggregate the aligned data in frequency, binning by a factor of 64 to produce 52 channels, in order to give sufficient signal that the pulsar profiles are clearly visible across the whole band. In Figures \ref{fig:7}--\ref{fig:10} we present the data for PSRs~J1359--6038 and J1056--6258 as waterfall plots, with phase on the x-axis and decreasing frequency on the y-axis. The plots are truncated in phase to show only the on-pulse region. The shading indicates the magnitude of the variable plotted, with total intensity shown in greyscale and the PA in colour. The intensity has been normalized by dividing each frequency channel by the summed flux in that channel. This is to emphasize pulse structure over intrinsic intensity, so that features of the profile at higher frequencies are not lost due to the spectral index of the emission. The figures displaying PA use a cyclic colour scheme. For each pulsar, the data are shown aligned both using the catalogue values for RM and DM, and using the corrected values calculated in this work.

These figures show that pulsar emission, both total intensity and polarization, varies smoothly with frequency across a broad band. As a result, it is possible to align the pulse profiles consistently across frequency. We recommend the style of figure produced here as an informative way to display broadband polarimetric data. The visualization technique may also be used to display Stokes V, which we do not present here but will address in a future publication.
For PSR~J1359--6038 it is clear that the catalogue DM has been selected to align the intensity peak across frequency (Fig. \ref{fig:8}, the white line indicates the average position of the intensity peak). Since this pulsar is scattered, which shifts power in the profile to later phases, this means that the peak of the unscattered profile is not aligned at the lowest frequencies. We performed scattering fits to the ten lowest frequency channels using the model of \cite{Geyer2016} to identify the positions of the unscattered profile peaks. These are marked with white dots on Fig. \ref{fig:8}.
The catalogue values of RM and DM do not align the polarization data, as can be seen in Fig. \ref{fig:7}. Using our method to align the PA shows a clear consistency of the PA shape from high to low frequency, broken only when the effects of scattering become prevalent below around 1300~MHz. This results in a noticeable misalignment in Stokes I (Fig. \ref{fig:8}), but one that is plausibly attributable to A/R, as described in section \ref{sec:RMDM}. The results for this pulsar show that it is possible to find an alignment of pulsar data using only corrections to RM and DM. This allows us to relate pulsar emission features at different frequencies and tie them to magnetospheric emission properties, independent of the effects of interstellar scattering.

Moving on to PSR~J1056--6258, it is clear that the catalogue values of RM and DM do not successfully align either the PA (Fig. \ref{fig:9}) or the total power (Fig. \ref{fig:10}), highlighting the problems that arise when trying to interpret pulsar radio emission using only discrete observations with narrow bandwidths. However, we are able to obtain a successful alignment of the PA which reveals a very consistent PA profile shape across the entire frequency band (Fig. \ref{fig:9}). At the highest frequencies there is some evidence of the orthogonal jump developing on the left of the profile (marked in dark purple on the figure), but with a low SNR due to its being at the edge of the pulse. The jump is found to be more obvious when a greater number of frequency channels are aggregated to build up SNR. The alignment of the PA results in a clear misalignment of the intensity profile (Fig. \ref{fig:10}), which, as for PSR~J1359--6038, can be attributed to the effect of A/R. It should further be noted that the intensity profile structure appears to be very consistent for this pulsar, with three components easily identifiable across the whole band. From this pulsar we learn that, although a flatter PA geometry makes it harder to constrain the alignment, it is still possible and important to do so. Furthermore, we highlight the necessity of fitting for both the DM and RM simultaneously.

\cite{Karastergiou2007} developed an empirical model of radio pulsar beams to describe the diverse population of observed pulsar profiles. This model proposed different emission height structures for pulsars with different values of $\dot{E}$. They suggested that pulsars with low $\dot{E}$, such as PSR~J1056--6258 (see Table \ref{tab:2}), emitted over a broad range of emission heights, whereas those with high $\dot{E}$, like J1359--6038, would have their emission restricted to a much narrower range. Our calculations of emission height ranges for these two pulsars are in accordance with this idea, without providing further constraints.

\section{Concluding remarks}
\label{sec:conc}

Pulsar polarization is a powerful diagnostic of the radio emission mechanism and geometry. In this paper, we have used new polarimetric data covering a continuous bandwidth of 700--4000 MHz to address questions regarding alignment across the band and the intrinsic frequency evolution of pulsar polarization. The key points of this work are as follows: 
\begin{enumerate}
    \item We developed a novel technique to align the rotational phase across the band with respect to the PA profile shape, motivated by its geometric origins. The PA profiles were restricted to shift only by corrections to the applied RM and DM. 
    \item We verified this technique through simulations, which we also used to develop our understanding of how to interpret the results. We found that alignment accuracy is dependent on PA curvature and that, despite not following the same squared relationship to frequency as dispersion and Faraday rotation, PA shifts due to frequency-dependent emission height can be absorbed into the calculated $\Delta$DM and $\Delta$RM corrections. The $\Delta$DM correction then results in a frequency dependent offset in the total power profiles. 
    \item Applying our methodology to PSRs J1056--6258 and J1359--6038 we found that it is possible to align the PA across frequency when constrained to using only the DM and RM to do so. The best fit values of RM and DM are covariant.
    \item We applied our corrections to the DM and RM and created visualizations of the aligned data to show how polarization evolves across frequency.
    \item For both pulsars we found that alignment of the PA profile leads to misalignment of the total intensity, mirroring what is seen in the simulations when variable emission height with frequency is introduced.
    \item With the assumption that for both pulsars this result is attributable to A/R, we estimated the height ranges across this frequency band. We note that the complexity of the frequency dependence of pulsar polarization, with multiple observational effects that are poorly explained by current theory, means that alternative origins are possible for the misalignment seen. Nevertheless, our predicted height ranges are compatible with those published in the literature, suggesting that A/R is a plausible explanation for this observation.
    \item \cite{Karastergiou2007} postulated that the emitting height range evolves as a function of $\dot{E}$, with high $\dot{E}$ pulsars emitting from a narrower range of heights than low $\dot{E}$ pulsars. Our results are in agreement. 
\end{enumerate}

These data show great promise for directing our developing understanding of the pulsar radio beam structure and emission mechanism. In this context, the work presented here will be followed by a comprehensive study of the broadband polarimetric properties of a large number of pulsars.
Applying our methodology more widely will allow us to test the assumptions and conclusions presented here in the presence of a greater variety of complex polarization effects, focusing also on the frequency dependence of circular polarization. 
We will further concentrate on questions of beam geometry and emission heights, rotational phase, and correlations with physical parameters such as $\dot{E}$ across a broader population.

\section*{Acknowledgements}
LO acknowledges funding from the Science and Technology Facilities Council (STFC) Grant Code ST/R505006/1.
The Parkes radio telescope is part of the Australia Telescope National Facility which is funded by the Australian Government for operation as a National Facility managed by CSIRO.

\bibliographystyle{mnras}
\bibliography{Polarization_paper}

% Don't change these lines
\bsp	% typesetting comment
\label{lastpage}

\end{document}